\documentstyle[11pt,newpasp,twoside,epsf]{article}
\def\edcomment#1{\iffalse\marginpar{\raggedright\sl#1\/}\else\relax\fi}
\reversemarginpar

\begin{document}
\title{Abundance of Terrestrial Planets by Microlensing}
 \author{Philip Yock}
\affil{Faculty of Science, University of Auckland, Auckland, New Zealand}

\begin{abstract}
Terrestrial planets may be detected using the gravitational microlensing technique. This was demonstrated in the high magnification event MACHO-98-BLG-35. Observing strategies aimed at measuring the abundance of terrestrial planets are discussed, using both existing telescopes and planned telescopes.   
\end{abstract}

\section{Introduction}

Significant advances have been made in recent years in the study of extra-solar planets using the radial velocity technique and, more recently, the transit technique (Mayor, these proceedings; Fischer, these proceedings). Approximately thirty Jupiter-mass planets have been detected by the radial velocity technique. Initial indications from transit measurements indicate they are gas giants similar to Jupiter. However, they have been detected in eccentric orbits at $>$ 0.2 AU or circular orbits at $<$ 0.2 AU, quite dissimilar to Jupiter's orbit. These surprising findings will assist our understanding of planetary systems and planetary formation. 

The gravitational microlensing technique complements the above studies because it is sensitive to giant and terrestrial planets at orbital radii of a few AU. In this technique, perturbations of standard microlensing caused by planets are searched for. Two versions of the technique have been utilised to date. These involve large perturbations of low-magnification events, and small perturbations of high-magnification events, respectively. Most searches have been made in the galactic bulge where the observed rate of microlensing is maximal.  

In the following sections strategies for optimising the detection rate of planets by microlensing, especially low-mass terrestrial planets, are discussed. The ultimate goal is to determine their abundances. To illustrate the technique, a high magnification event is described in the following section. Other examples of both high and low magnification events are described by Gaudi in these proceedings.       

\section{High-magnification event MACHO-98-BLG-35}               

The event MACHO-98-BLG-35 has been described elsewhere in some detail (Rhie et al. 2000). The peak magnification was $\sim$80, the highest observed to date. Here just those features of the event that demonstrate the sensitivity of the microlensing technique to the detection of low-mass planets are described. 

The peak of MACHO-98-BLG-35 was monitored fairly extensively from Australia and New Zealand by the MPS and MOA groups. Their light curves are reproduced in Figure 1. Some features are noteworthy. The light curves are incomplete. Nearly half of the peak was not monitored. Second, the photometry is accurate to $\sim$1.5\% only. This was achieved with the standard DoPHOT programme.

\begin{figure}
\plottwo{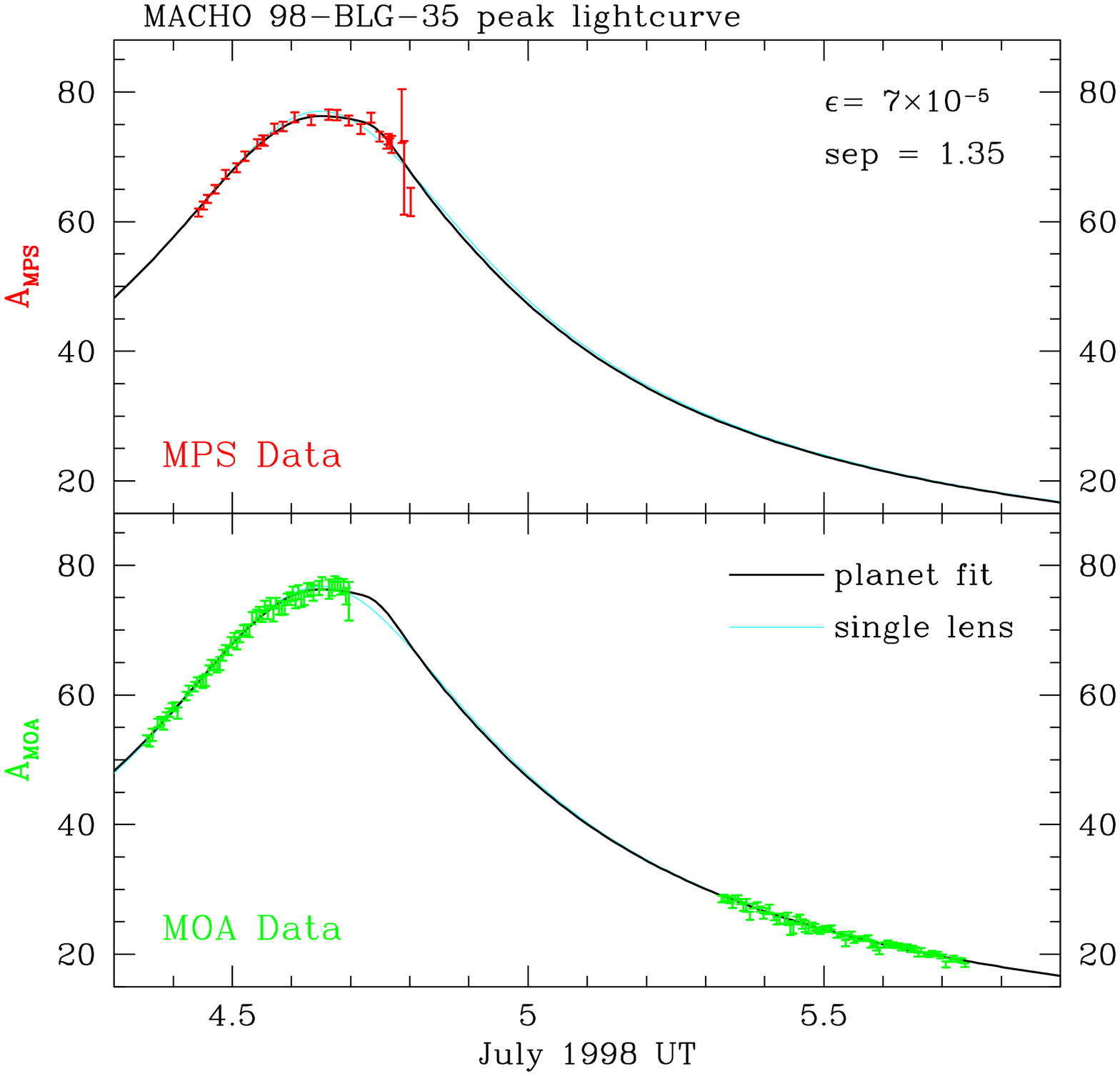}{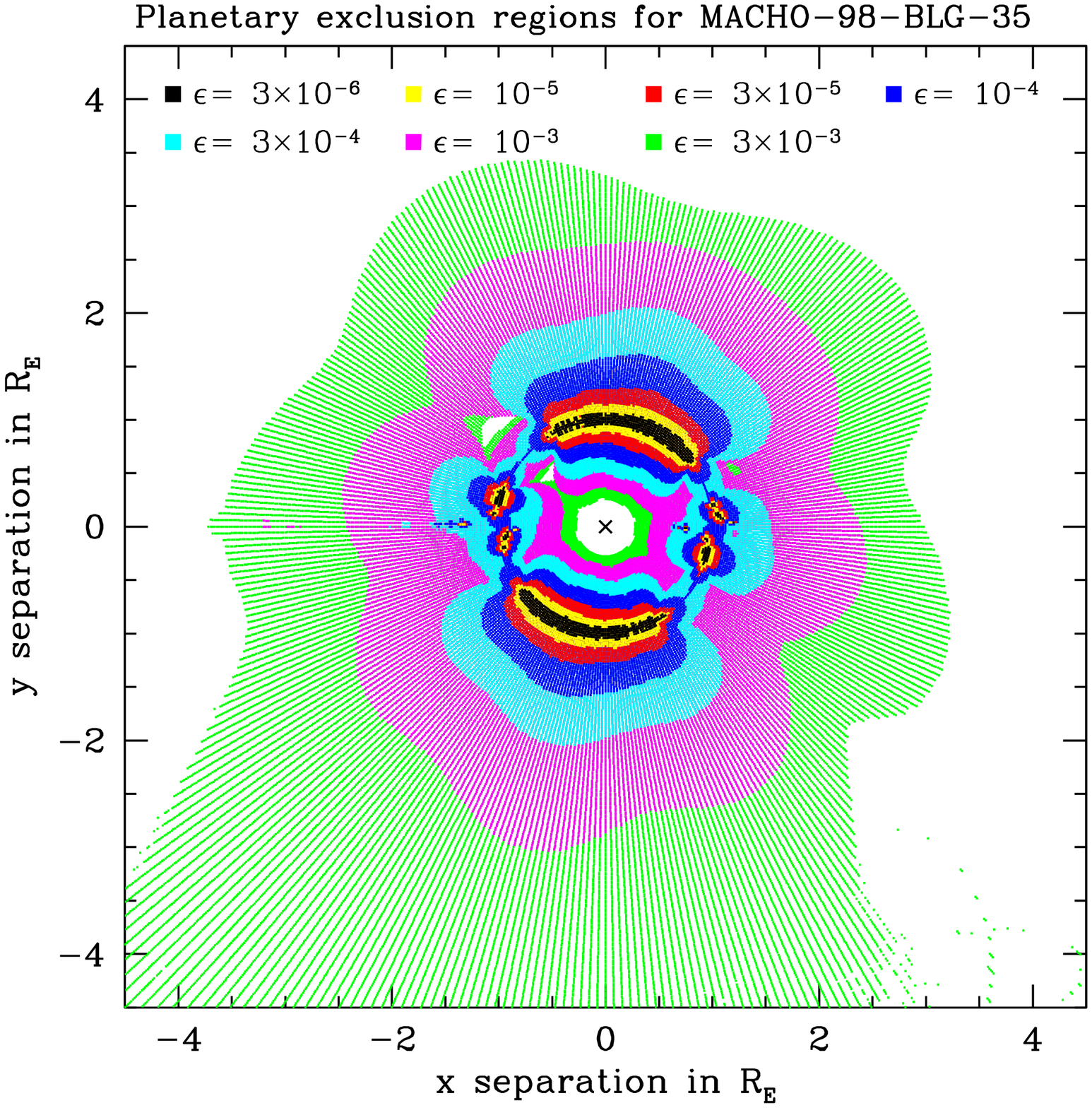}
\caption{Light curves and 'snapshot' of the planetary system in the microlens event MACHO-98-BLG-35 from Rhie et al. (2000). The pale light curves are the best fits to the data for a lens without a planet, and the heavy curves are the best fits for a lens with a single planet with mass fraction $7\times10^{-5}$ and orbit radius 1.35 Einstein radii. The right panel shows the exclusion regions for planets with various mass fractions. The two unshaded regions at about 10 o'clock designate the possible locations of a planet at the time of the observations.}
\end{figure}

Despite the above limitations, significant information on the presence of planets orbiting the lens star for this event was obtained. This is illustrated in the right-hand panel of Figure 1. This is a two-dimensional 'snap shot' of the lens's planetary system at the time of the event. It shows exclusion regions (in projection) at the 6.3$\sigma$ level of confidence for planets with masses ranging from 3$\times10^{-6}$ to 3$\times10^{-3}$ times that of the lens. Assuming a most likely lens mass $\sim0.3M_{\odot}$, this range corresponds to 0.3 Earth-mass to Jupiter-mass. The exclusion region for the lowest mass is the dark, nearly circular tube at the Einstein radius ($r=1R_E$). Here $R_E$ is $\sim$ 0.6-2.8 AU. The exclusion regions for heavier planets are successively thicker tubes surrounding this. The two unshaded regions just inside and outside the Einstein radius at about 10 o'clock are two possible locations (in projection) at the time of the observations of a planet with fractional mass $4\times10^{-5}$ to $2\times10^{-4}$. The evidence for this planet is at the 4.5$\sigma$ level of confidence. 

These results indicate both the potential and the limitations of the microlensing technique. Only a 'snap-shot' is obtained. There is essentially no chance for follow-up measurements. Also, the photometry does not uniquely determine all the parameters of the system. Balanced against these limitations is the demonstrated potential of the technique to detect low-mass planets. A significant volume surrounding the lens-star of MACHO-98-BLG-35 was shown to be devoid of Earth-like planets.

Improvements in the precision of the above results for MACHO-98-BLG-35 should be possible with the inclusion of more data and better photometry. Data by the PLANET group are available (J. Greenhill, private communication) but these have not been incorporated as yet. Also, a re-analysis of the data using subtraction photometry may yield higher precision (Bond, these proceedings). A 'gigablaster' similar to those used for human genome sequencing (Claverie 2000; Gee 2000) is being used for this purpose.                     

\section{Optimising the detection rate of terrestrial planets by microlensing with existing telescopes}

The microlensing technique may be understood physically as follows. In high-magnification events, good alignment of the lens and source stars occurs. This  causes a circular, or near-circular, image to form around the lens. If a planet is present in such an event near the (projected) Einstein radius it will perturb the image. The perturbation will be small, however, because most of the image will lie far from the planet and remain unperturbed. This is the case illustrated in the left panel of Figure 1. In low-magnification events the image consists of two, short, rotating arcs either side of the lens star. If either of these approaches a planet (in projection) during an event, a relatively large perturbation of the light curve occurs. However, in most cases, neither image will approach a planet. Moreover, the relatively low light-level of low-magnification events results in photometry of poorer accuracy. In both high and low magnification events, the Einstein radius is typically $\sim$ a few AU. This enables planets at these orbital radii to be detected.
 
To date, the most fruitful microlensing events for planetary searches have been the high magnification ones. This is primarily because of the high probability of detectable planetary perturbations occurring in the peaks of these events, if planets are present. This was first predicted by Griest and Safizadeh (1998). Also, the short durations of high-magnification peaks permit them to be sampled densely without undue difficulty. This enhances the probability of detecting low-mass planets for which the planetary perturbation is only brief. Additionally, the high light-levels attained in high magnification events enable good photometry to be performed with small telescopes.   

If no drastic changes occur in the observing strategies of the current microlensing groups in the next few years, it may be anticipated that the current results will be extended somewhat, but not greatly. A tighter constraint on the abundance of giant planets at a few AU than that reported by Gaudi in these proceedings can be expected, and possibly some detections. Also, further constraints and/or results on the presence of terrestrial planets can be expected from events like MACHO-98-BLG-35. In the same period, the abundance of giant planets at a few AU is likely to be determined by the radial velocity technique.  

Further information on terrestrial planets would be obtained if more telescopes were devoted to the microlensing technique. Currently there are some ten 1-m class telescopes in Chile, South Africa and Australasia that are being used for microlensing. If all these were dedicated to the relentless observation of the peaks of high magnification events, thus minimising gaps in light curves caused by inclement weather, tighter constraints and/or results on the presence of terrstrial planets would be obtained. If, furthermore, rapid on-line analysis was employed by more survey groups to alert on high magnification events in progress, the detection rate would be additionally enhanced. A first rough indication of whether or not terrestrial planets are common could then be possible.  

\section{Measuring the abundance of terrestrial planets by microlensing with new telescopes}

High-quality planetary abundance measurements could be made with a single 2-m class wide-field telescope monitoring a few square degrees of the galactic bulge at a sampling rate of a few observations per hour from the Antarctic or from space (Sahu 1998; Yock 2000; Bennett \& Rhie 2000). The rate of detection of microlensing events with such an instrument would significantly exceed the rate currently being detected, especially if a near infrared passband was used. This would enable the galactic centre to be monitored where the microlensing rate is expected to be highest (Gould 1995). Additionally, the high sampling rate would result in both high and low magnification events being densely sampled, increasing the detection rate of low-mass planets. A simple scaling of the results obtained to date suggests that useful planetary abundance measurements could be made in a few years from the Antarctic. Specific predictions require further site testing to be carried out at the Antarctic. This is currently in progress at the high-altitude site known as Dome-C (Burton 1998). Probably the ideal is the proposal that was made recently by Bennett and Rhie (2000) for a space-based $\sim$ 1.5m diffraction limited wide-field telescope known as GEST. This would have the capability to detect $\sim$ 100 Earth-like planets in 2.5 years if such planets are common.              

\section{References}
Bennett, D.P. \& Rhie, S.H. 2000, in Disks, Planetesimals and Planets, ASP Conference Series, ed. F. Garzsn, Carlos Eiroa, Dolf de Winter \& T.J. Mahoney (San Francisco: ASP)\\  
Burton, M.G., 1998, in Astrophysics from Antarctica, ASP Conf. Series 141, ed. G. Novak \& R. Landsberg (San Francisco: ASP), p.3\\
Claverie, J-M. 2000, Nature 403, 12\\
Gee, H. 2000, Nature 404, 214\\
Griest, K. \& Safizadeh, N. 1998, ApJ, 500, 37\\
Gould, A. 1995, ApJ, 446, L71\\
Rhie, S.H. et al. 2000, ApJ, 532 \\
Sahu, K., 1998, in Astrophysics from Antarctica, ASP Conf. Series 141, ed. G. Novak \& R. Landsberg (San Francisco: ASP), p.179\\
Yock, P. 2000, PASA, 17\\

\end{document}